# DEVELOPMENT OF VIDEO FRAME ENHANCEMENT TECHNIQUE USING PIXEL INTENSITY ANALYSIS

**H. A. Abdulkareem, A. M. S. Tekanyi, I. Yau and B. O. Sadiq**
**ABU FM Radio & BBC Studio, Ahmadu Bello University, Zaria, Nigeria**
ha2zx@yahoo.com; amtekanyi@abu.edu.ng; isiyakuyau@abu.edu.ng; hamisu99@yahoo.com

***ABSTRACT:*** *This paper developed a brightness enhancement technique for video frame pixel intensity improvement. Frames extracted from the six sample video data used in this work were stored in the form of images in a buffer. Noise was added to the extracted image frames to vary the intensity of their pixels so that the pixel values of the noisy images differ from their true values in order to determine the efficiency of the developed technique. Simulation results showed that, the developed technique was efficient with an improved pixel intensity and histogram distribution. The Peak to Signal Noise Ratio evaluation showed that the efficiency of the developed technique for both grayscale and coloured video frames were improved by PSNR of 12.45%, 16.32%, 27.57% and 19.83% over the grey level colour (black and white) for the NAELS1.avi, NAELS2.avi, NTA1.avi and NTA2.avi respectively. Also, a percentage improvement of 28.93% and 31.68% were obtained for the coloured image over the grey level image for Akiyo.avi and Forman.avi benchmark video frame, respectively.*

## 1. INTRODUCTION

Digital images can be considered as a large array of discrete dots, each of which has a brightness associated with it. These dots are called picture elements or simply pixels (Pandey *et al.,* 2015). Pixels are the smallest discrete component of an image. These pixels are often corrupted with noises which are inherent in the image. Therefore, before further processing of an image or video data, it is necessary to remove noise (Boon and Guleryuz, 2006), (Siddavatam *et al.,* 2011), (Roopashree et al., 2012). Image pre-processing is the term used for operations on images at the lowest level of abstraction. These operations do not increase image information content but they decrease its entropy information measure (Fan and Jin, 2010), (Kamboj and Rani, 2013), (Garg and Kumar, 2012). Neighbouring pixels corresponding to one real object have the same or similar brightness value (Li *et al.,* 2010). If a distorted pixel can be picked out from the image, it can be restored as an average value of neighbouring pixels (Li *et al.*, 2011). Image pre-processing methods can be classified into categories according to the size of the pixel neighbourhood that is used for the calculation of new pixel brightness (Padanathi *et al.,* 2012),. (Brochier *et al.,* 2015). In literatures, images are broadly classified into two types which are the grey scale image and the coloured images (Verima and Ali, 2015). The grey scale images are typically 2D images and the coloured images are the 3D images. The primary reason for pre-processing these types of images is to remove noise (Ballabeni *et al.,* 2015). A number of researchers have explored new and improved ways of enhancing and restoring video quality such as the works of (Zhang *et al.,* 2013), (Ghailke and Ganorker, 2013). The authors in (Zhang *et al.,* 2013) developed a fast-recursive algorithm to estimate the decoder-side distortion of each frame in the presence of packet loss. The algorithm operated at block level and considered the impacts of different intra prediction modes. An optimization problem was formulated to minimize the decoder-side distortion by allocating a given channel coding redundancy among a group of frames. Various techniques were introduced to speed up the algorithm without sacrificing too much accuracy in order to meet the hardware and real-time constraints of the system. As a result, the developed scheme could run on a real-time embedded video conferencing system with resolution of up to 1024×576 pixels, 30 frames per second (fps) and 4 megabits per second (Mbps). Luminosity conserving and contrast enhancing histogram equalization method for color images was developed by (Gwangil, 2014). All RGB images were transformed into different color spaces and particular channels were applied to the histogram equalization process. The final result showed that HSV color space yielded favourable results in MSE by giving good luminosity conserving ability. It is always established that HSV color space is characterized with a lot of drawback, some of which are freezing of color as a result of oscillation created by chips in the camera whose end result is noise. (Vishwakarma and Mishra, 2012), (Ghadke and Ganorker, 2013) critically reviewed the color image enhancement techniques. Their research work revealed that histogram equalization cannot preserve the brightness and color of the original image. In view of these, this paper aims to develop a brightness enhancement technique for video frame pixel improvement based on pixel intensity analysis to improve the quality of the video frame.

## 2. METHODOLOGY

The step by step procedure adopted in this research to develop, a novel image brightness enhancement technique based on image quality is highlighted as follows:

2.1 Acquisition of video data on which the developed technique was implemented.
2.2 Elimination of hue and saturation whilst retaining luminance intensity of images.
2.3 Noising and filtering of images.
2.4 Development of the achieved brightness enhancement technique

### 2.1 Video Acquisition

A total of four acquired video data and two benchmark video data were used in order to efficiently determine the performance of the developed technique under different conditions. The sample of these video frames is presented in Figure 1.

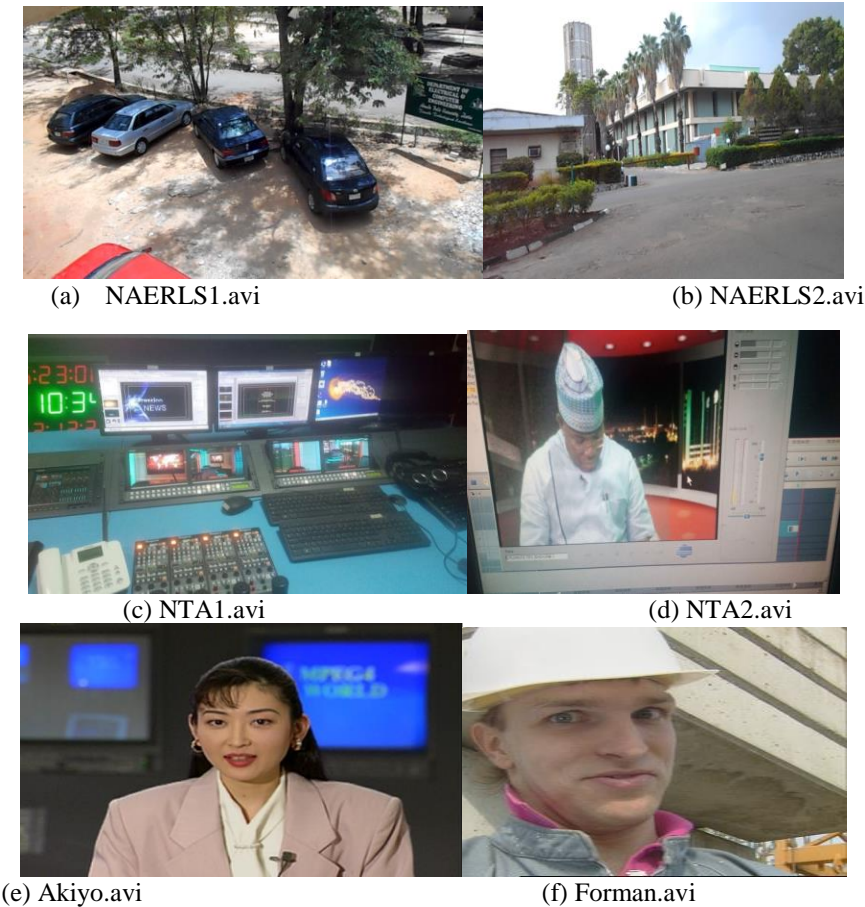

(a) NAERLS1.avi (b) NAERLS2.avi
(c) NTA1.avi (d) NTA2.avi
(e) Akiyo.avi (f) Forman.avi

Figure 1: Frames of Sampled Videos and Benchmark Videos Frames

From Figure1, the first two sample videos from NAERLS (NAERLS1.avi and NAERLS2.avi) were obtained using the video camera system of NAERLS. Similarly, the two sample videos from NTA (NTA1.avi and NTA2.avi) were obtained using the video camera system of NTA, Abuja Broadcasting Station and the last two benchmark videos (Akiyo. avi and Forman avi) were also obtained from the standard image database (VintaSoftImaging.Net). MATLAB R2015a image processing toolbox was used on individual video images obtained and represented in Figure 1 to achieve the information on each video as shown in Table1.

Table1: Simulation Result Sample Video Data

| SN | File Names | File Size (*.avi) | Number of file Frames |
|----|------------|-------------------|------------------------|
| 1  | NAERLS1.avi | 18.1Mb | 157 |
| 2  | NAERLS2.avi | 10.3Mb | 155 |
| 3  | NTA1.avi | 9.6Mb | 152 |
| 4  | NTA2.avi | 11.2Mb | 200 |
| 5  | Akiyo.avi | 11Mb | 300 |
| 6  | Foreman.avi | 7.25Mb | 100 |

It is noteworthy that, videos are frames of dynamically changing images and images are usually derived from video cameras in the form video frames. Therefore, the video data given in Table 1 were initially converted into frame of static images and cropped for easy processing and analysis.

**2.2 Eliminate Hue and Saturation and Retain Luminance Intensity**

The next stage in the pre-processing was to extract the luminance information from the true colour image by removing the hue and saturation. In order to achieve this, conversion formula was employed as follows (Keith, 2005):

$$Intensity = a \times (red) + b \times (green) + c \times (blue) \quad (1)$$

where a, b, and c are the hue weighted average of the images Red, Green, and Blue (RGB) channels. Note; the fundamental basic luminance (brightness) equation is given as
Y = 0.299R + 0.587G + 0.114B and the sum of the three coefficients of red, green and blue is 1 (Keith, 2005). These individual values of channels are usually greater than zero, depending on the colour content of the original image. These coefficients were adjusted based on sensitivity to colours until appropriate luminance intensity was obtained. The constituent colours of the image were extracted and the image intensity given in equation (1) was transformed into equation (2): (Keith, 2005)

$$Intensity = a \times (1,:,:) + b \times (:,2,:) + c \times (:,:,3) \quad (2)$$

From equation (2), the first matrix is with only the coefficient multiplying the Red, coefficient of both Blue and Green represented in hidden does not exist. The same procedure as in the second and third matrix where the Blue and Green exist.

**2.3 Noising and Filtering**

Noise was applied to the intensity image given in equation (2). Four different types of noise (Gaussian, Poisson, salt& pepper, and speckle) were implemented. The essence of this introduction of noise was to randomly vary the intensity of the image such that the pixel values show different values from its true values. By so doing, the actual image was separated from its background. Since the image intensity ranges from 0 to 1, an addition of parameter (*d*) was used to control the level of the noise added to the image. This was implemented using the inbuilt image processing toolbox in MATLAB R2015a and its snippet is given as follows:

```
IM=IMAGE;

IN1=imnoise(IM,'salt & pepper',d);

IN2=imnoise(IM,'poison',d);

IN3=imnoise(IM,'gaussian',d);

IN4=imnoise(IM,'speckle',d);
```

The Noisy Image (*IN*) from this operation was smoothed using the median filter and hybrid median filter which are two of the most widely used filtering technique. The median filtering was done by selecting the median value from the neighborhood as the output of each pixel (Sadiq and Sani, 2015). The inbuilt image processing toolbox in MATLAB R2015a was used for its implementation as follows:

```
IMF1=medfilt2 (IN, [K,L]);

IMF2=hmf (IN,[K,L]);
```

where, K and L are the selected neighborhood operation of the image and IN is the noisy image obtained from the previous operation.

**2.4 Development of the achieved Brightness Enhancement Technique**

This research work used a different enhancement technique based on the following approach:

i.) At the first stage, the pixel values throughout the image was extracted and stored in a buffer (B). The dimensional sizes of the image were first determined and stored as X number of rows and Y number of columns. Therefore, from MATLAB toolbox R2015a, each pixel value extraction formula is given as:

$$B(k+1) = \sum_{k=0}^{N} \left( \overline{\frac{I(k)}{X \times Y}} \right) \qquad (3)$$

where:

*N* is the total number of possible intensity levels, which are usually 256 for 8-bit grey scale image. I is the input image, note; the product of the row and column, X and Y is the area of the frame. Also consider the buffer B (k +1) as an empty box where each frame is sampled into, and that sampling of the frame starts from k = 0 to k = N as can be seen from the flowchart in Figure 2. The bar is an indication of the whole columns of the matrix representing the frames sampled into an empty box.

Equation (3) only extracts part of the image that constitute the pixel information ii.) At the next stage, the pixel information from equation (3) is converted into a pixel matrix as follows using MATLAB code.

$$W = \sigma \times [1, (N+1)] \qquad (4)$$

Therefore, substituting equation (4) into equation (3) gives the following:

where $\sigma$ is a variable coefficient (variable, so starts from zero) and $B(l+1)$ is the updated picture information.

$$B(l+1) = \sum_{m=0}^{N+1} B(m) + \sum_{l=0}^{N+1} w(l) \qquad (5)$$

B(*m*) is the pixel and w(*l*) is the matrix defining each pixel.

iii.) In this stage, the elements of the update equation (5) were converted to integer values for further processing as follows:

$$R = \lfloor B(l+N) \times (N+1) + 0.5 \rfloor \qquad (6)$$

The addition of 0.5 in equation (6) was to ensure that the element of the image stored as R was always rounded up to 1 because this is when the maximum intensity of the image is achieved. The probability density function of brightness is always rounded to 1 (Ian *et al.,* 1995).

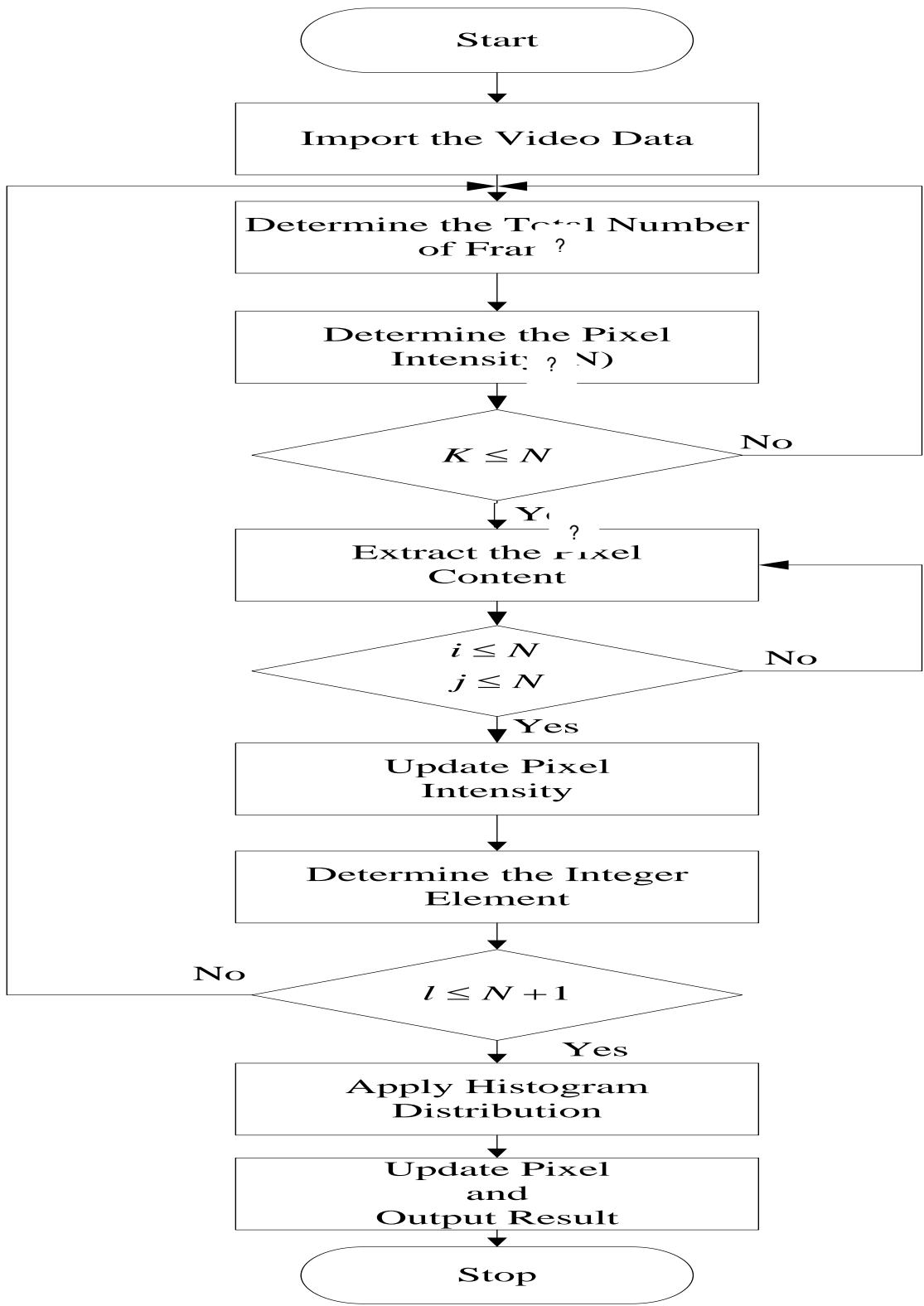

*Figure 2: Flowchart Implementation of Luminance Enhancement Technique*

The values of equation (3) that correspond to any of the values of R were determined and stored as *P* using the following:

$$P = \sum_{i}^{N+1} sum(B(find(R=i))) \quad (7)$$

where sum and find are operators in the MATLAB code. Equation (7) is the improved brightness or luminance of the employed method that enhanced the level of luminance intensity.

In the final stage, the pixel values of the original image were compared with R and the best from both images was kept. This was achieved using the following formula:

$$F\left(find(I=i)\right) = \sum_{i=0}^{N} R(i+1) \quad (8)$$

Note; the use of *find* is very necessary in equation (8), was to determine the linear indices corresponding to the nonzero element of the image.

The output from equation (8) was the improved brightness or luminance of the used image brightness enhancement technique. This technique is robust to any kind of image whether coloured or grey scale images these were evident in the result of the enhancement model.

## 3. RESULTS AND DISCUSSION

The acquired video samples were read and converted into image frames, which were resized to (176x144) which is equivalent to 43.3kB. Hue and saturation were eliminated to give images shown in Figure 3.

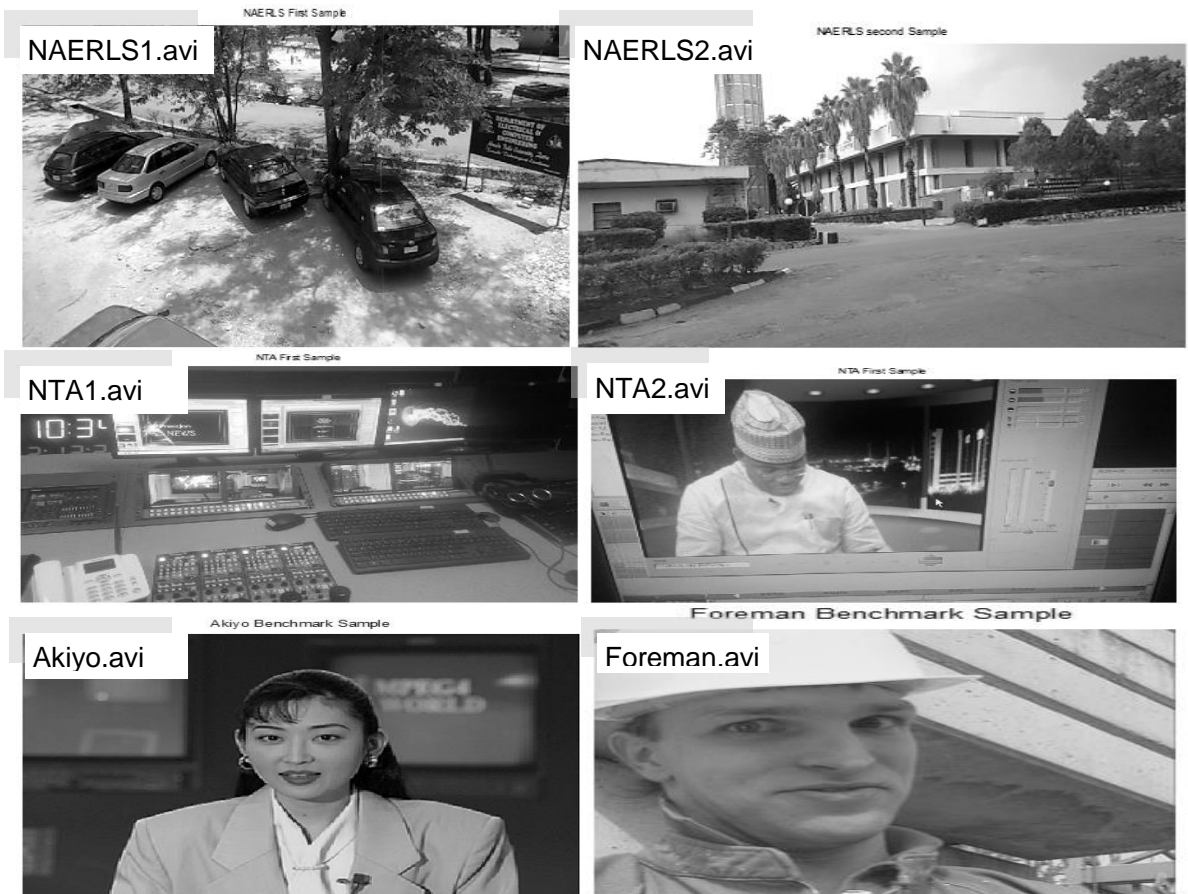

Figure 3: Luminance Intensity of Black and White Video Frames

In order to determine the efficiency of the luminance enhancement technique, the pixel values of the hue and saturation free image in Figure 3 were randomly varied by the introduction of noise (see its snippet in section 2.3), then filtered to smoothen the output. The filtered image was passed through the luminance enhancement technique and the result is shown Figure 4.

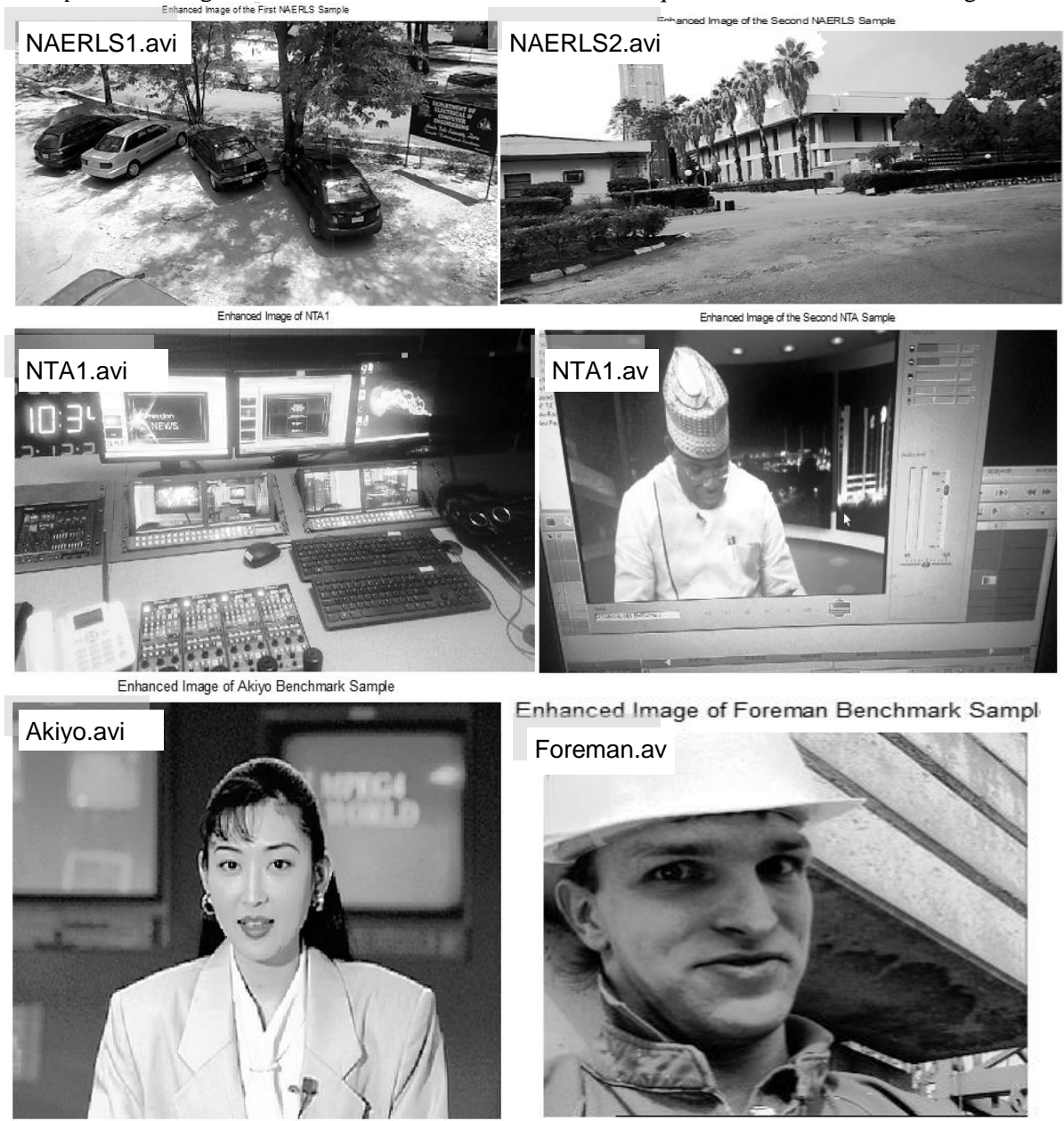

Figure 4: Luminance Intensity of Black and White Video Frames after Enhancement

The impact of the technique (luminance enhancement) used clearly shows that, each frame in Figure 4. appears to be brighter, sharper and rich in contrast when compared with their equivalent frames in Figure 3. In order to provide a clear justification for this statement, the histogram of each of the sampled video frames and the benchmark video frames were generated from the MATLAB stimulation environment as shown in Figure 5. This is the vectors representation of the image pixels at different intensities using histogram distribution.

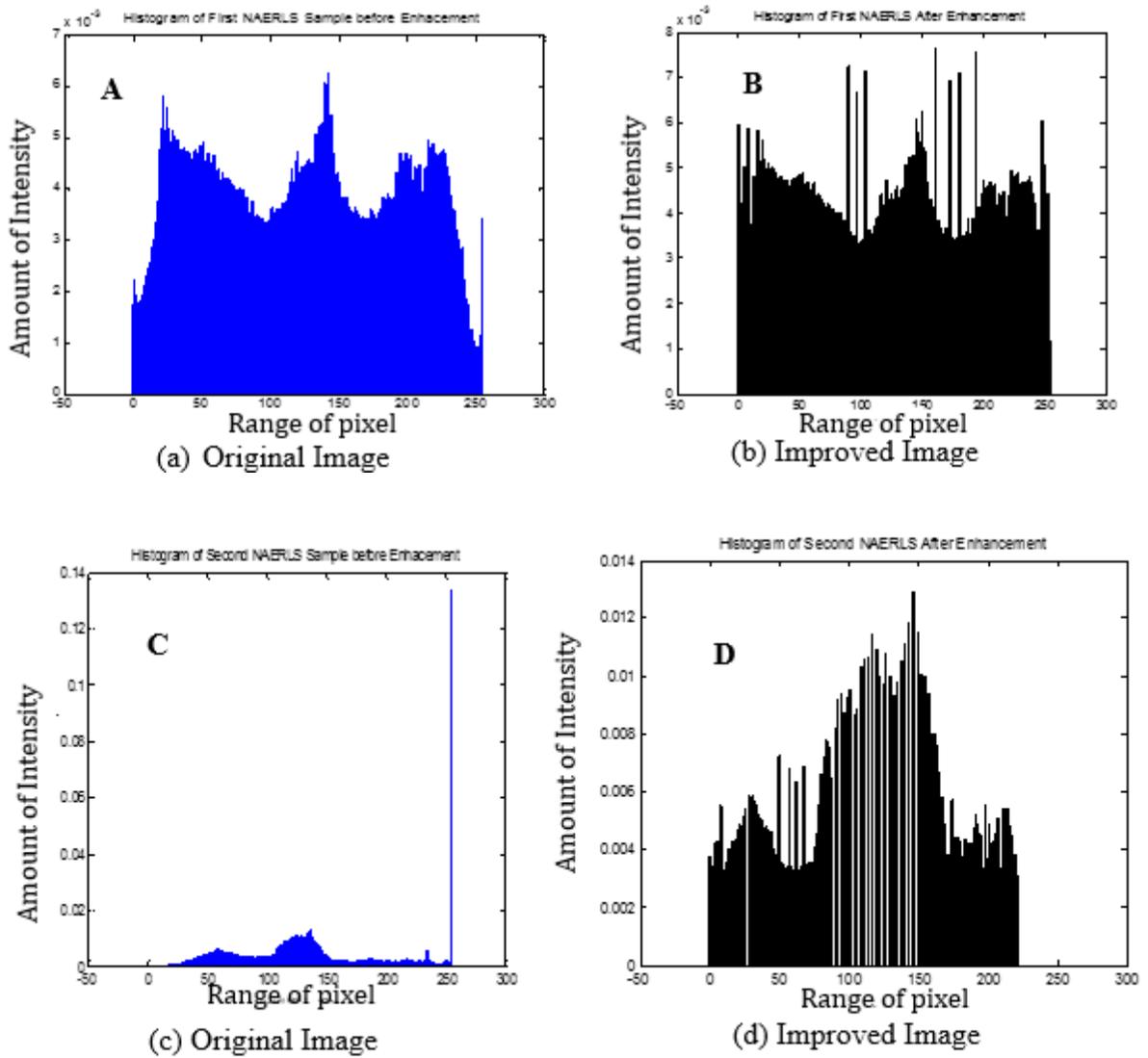

Figure 5: Histograms of Original and Improved NAERLS Sample Images

Figure 5 (a & b) show histogram representation of the NAERLS1 and NAERLS2 sampled image before the application of the luminance enhancement technique while Figure 5 (b & d) Show the histogram of the frames after enhancement. It should be observed that, the luminous intensity of the NAERLS1 frame (Figure 5 a) has been significantly improved and distributed across the pixel values in Figure 5 (b). Similarly, the luminous intensity of NAERLS2 sample frame given in Figure 5(c) is improved significantly across the pixel values of the frame in Figure 5(d). Note this is the improvement of the source original image in terms of the PSNR.

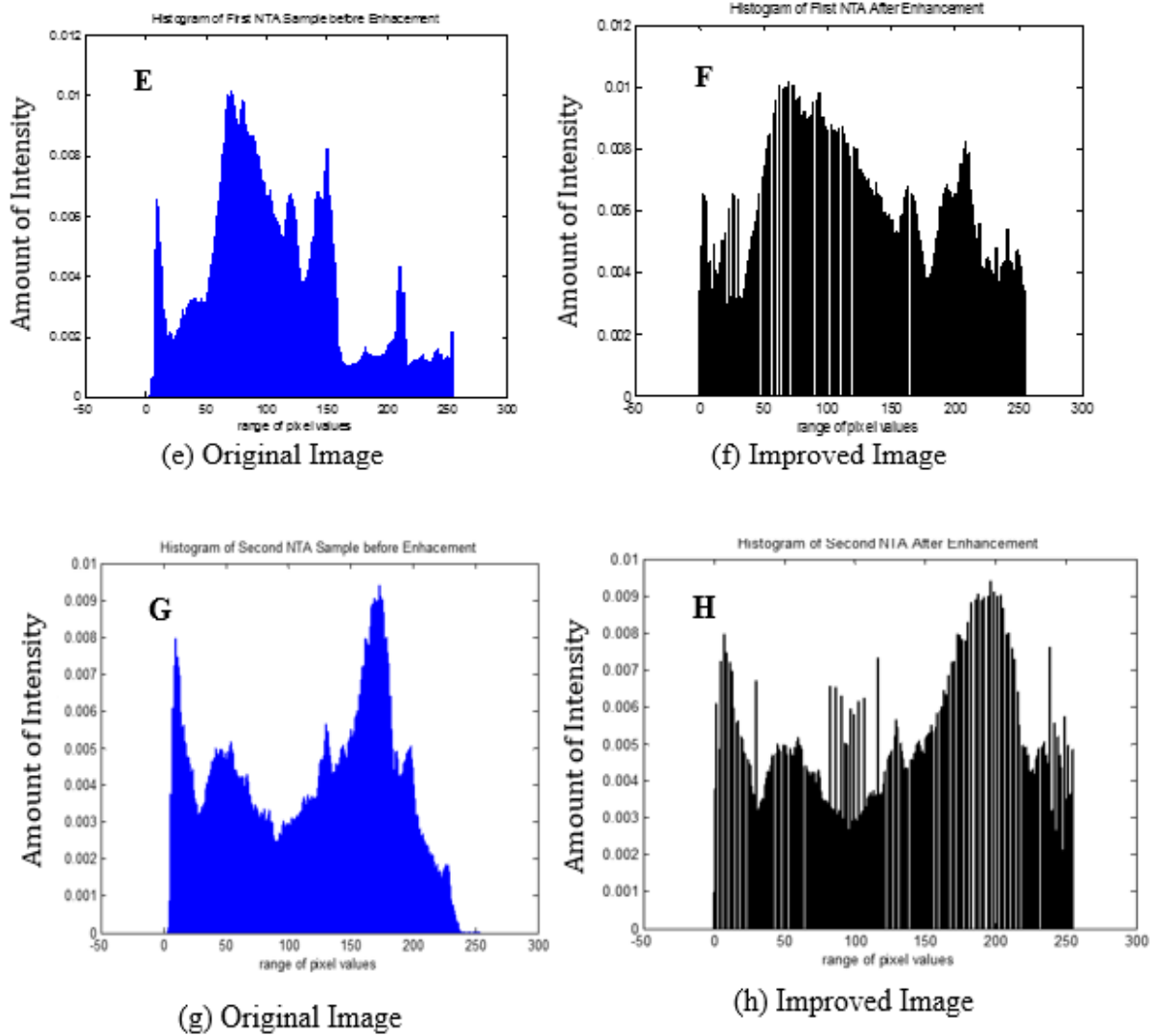

Figure 6: Histograms of Original and Improved NTA Sample Images

Figure 6 (e) and Figure 6 (f) show the respective histograms of the NTA1 sample video frames before and after luminance enhancement. Also, Figure 6 (g) and Figure 6 (h) show the histogram of the NTA2 sample video frame before and after the luminance enhancement It is observed from both Figure 6 (f) and Figure 6 (h) that the histograms of the frames from both cases have an improved luminous intensity with a much better distribution across the pixels values of each frame. This is an indication of how much of enhancement is done on the frames.

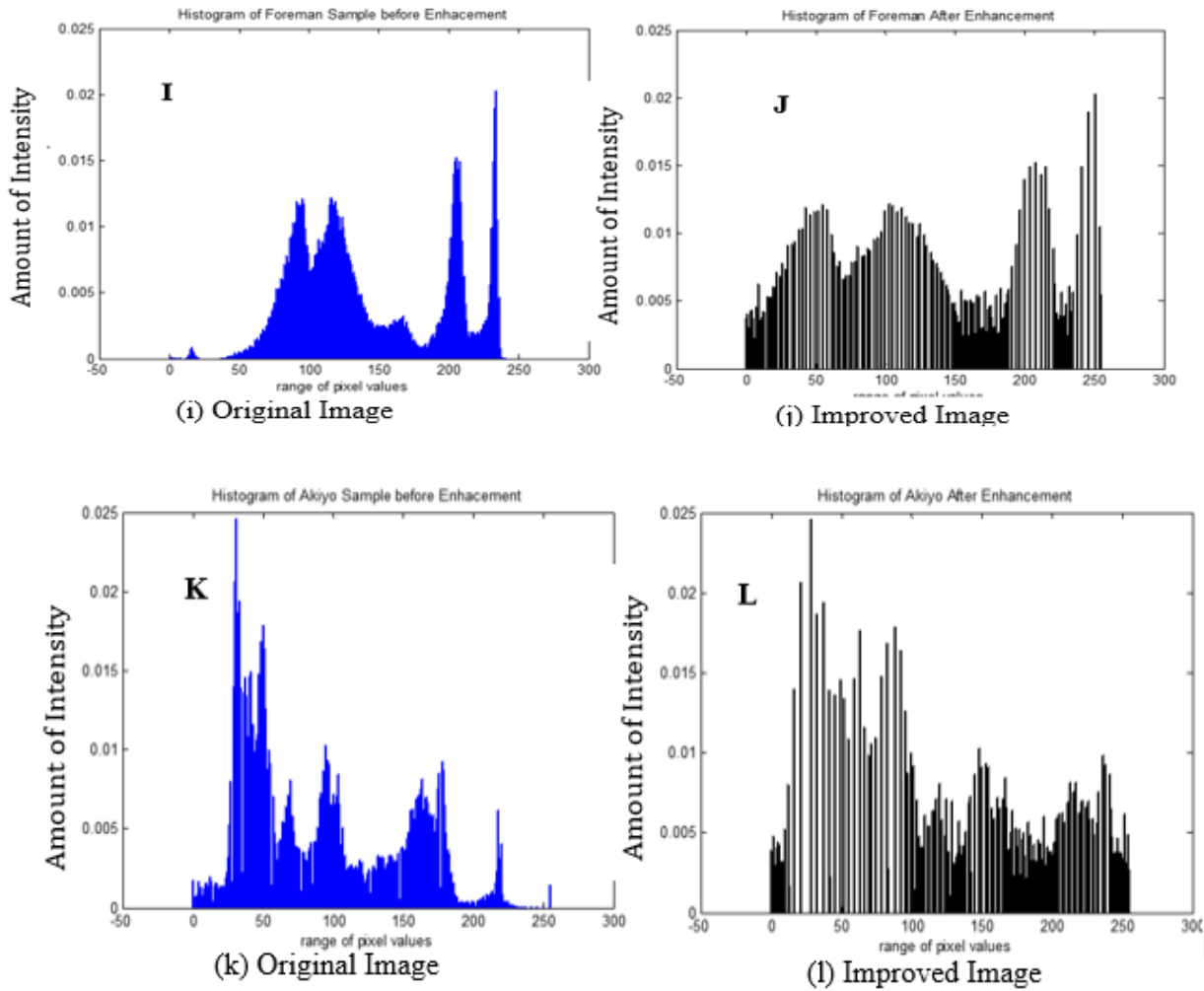

Figure 7: Histograms of Original and Improved Benchmark Sample Images

Figure 7(i) and Figure 7(j) show the respective histograms of the *Foreman.avi* benchmark video frames before and after luminance enhancement, while Figure 7(k) and Figure 7(l) show the histograms of the *Akiyo.avi* benchmark video frame. In both figures, it is observed that the intensity of the frames has improved throughout the pixels values with a much better distribution. This also, is an indication of how much enhancement is done on the benchmark frames. In order to evaluate the robustness of the luminance enhancement techniques presented in this report, the technique is applied to the coloured frame of all the six sampled videos. This is to determine how much of improvement can be done on the frames containing hue, saturation and luminance all together using the brightness (luminance) enhancement technique.

The enhancement technique was also applied to the coloured sampled frame of the two benchmark videos. The original and enhanced frames of Akiyo benchmark video, with their corresponding histograms are given in Figure 8.

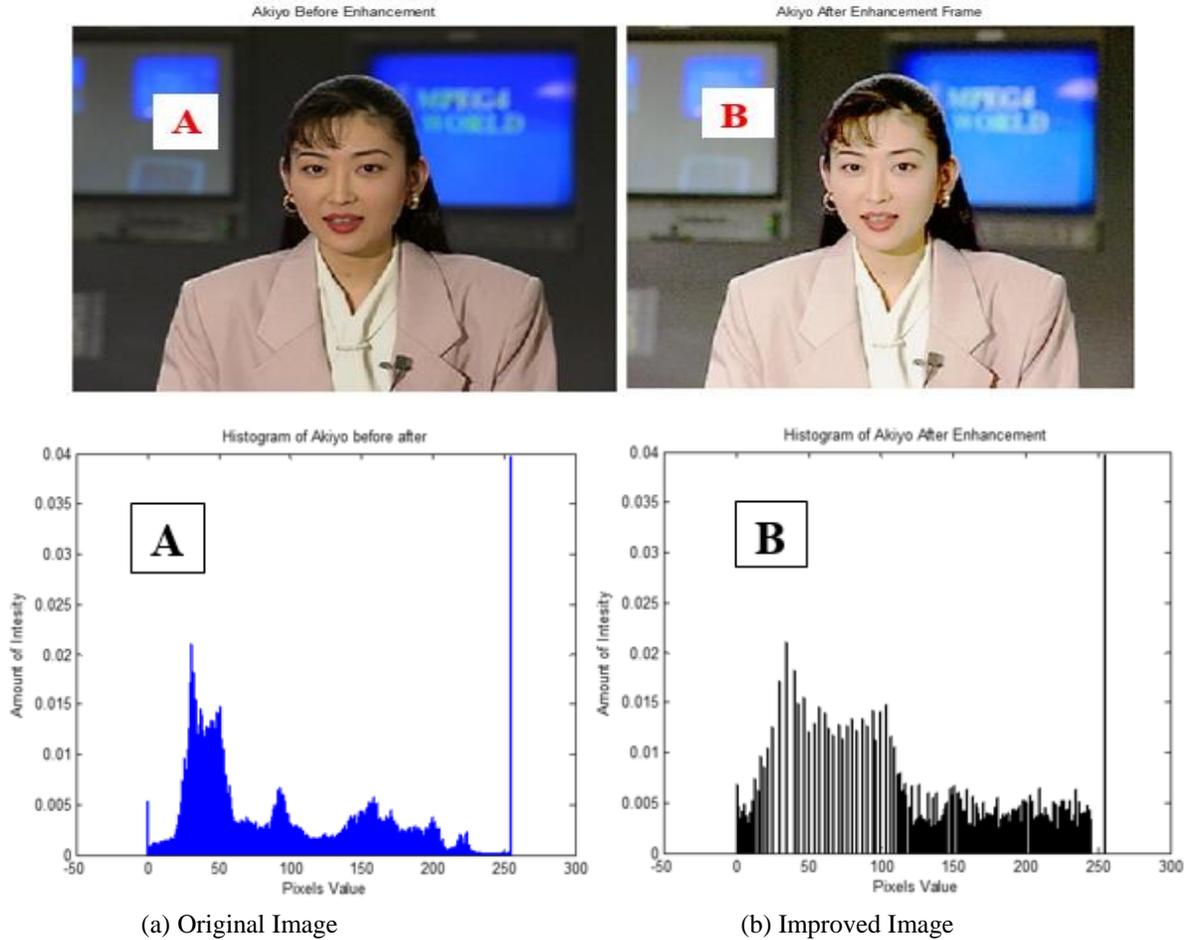

(a) Original Image  (b) Improved Image

Figure 8: Coloured Images and Histograms of Original and Improved sample of Akiyo

From Figure 8, it is observed that the histogram of the original *Akiyo.av* benchmark sample video frame of Figure 8(a) shows a much better improvement in Figure 8(b) after the application of the enhancement technique. Table 2 is the black and white (grey level) performance evaluation using Peak Signal-to-Noise Ratio.

### 3.1 Performance Evaluation using Peak Signal to Noise Ratio

In order to further evaluate the performance of this method, the Peak Signal-to-Noise Ratio (PSNR) is used to measure the quality of reconstructed sample frames of video signals.

Table 2: Simulation Result of Performance Evaluation using PSNR on Black and White Frames

| S/N | SAMPLE | SIZE | FRAME | PSNR |
|---|---|---|---|---|
| 1 | NAERLS1.avi | 18.1Mb | 157 | 31.95Db |
| 2 | NAERLS2.avi | 10.3Mb | 155 | 22.30dB |
| 3 | NTA1.avi | 9.6Mb | 152 | 17.71dB |
| 4 | NTA2.avi | 11.2Mb | 200 | 23.17dB |
| 5 | Akiyo.avi | 11Mb | 300 | 15.06dB |
| 6 | Foreman.avi | 7.25Mb | 100 | 19.17dB |

The PSNR results of the black and white sampled video frame signals and coloured sampled video frame signals given in Table 2 were obtained from Figure 4 and Figure 8, respectively. In order to make a justifiable conclusion, the PSNR of the brightness enhancement technique on the coloured sample frames were determined as shown in Table 3. Note all tables in this section were computed from MATHLAB simulation environment R 2014a.

Table.3 Simulation Result of Performance Evaluation using PSNR on Coloured Frames.

| S/N | SAMPLE | SIZE | FRAME | PSNR |
|---|---|---|---|---|
| 1 | NAERLS1.avi | 18.1Mb | 157 | 36.45dB |
| 2 | NAERLS2.avi | 10.3Mb | 155 | 26.65dB |
| 3 | NTA1.avi | 9.6Mb | 152 | 24.45dB |
| 4 | NTA2.avi | 11.2Mb | 200 | 28.90dB |
| 5 | Akiyo.avi | 11.0Mb | 300 | 21.19dB |
| 6 | Foreman.avi | 7.25Mb | 100 | 28.06dB |

The PSNR decibel values of Table 3 for the coloured frames show improvement when compared with their corresponding values in Table 2 for the hue and saturation free (black and white) frames. This indicates that, the brightness enhancement technique is more effective and efficient for coloured video than hue and saturation free frames. The bar chart in Figure 9 shows the amount of improvement on the coloured frames over the black and white frames.

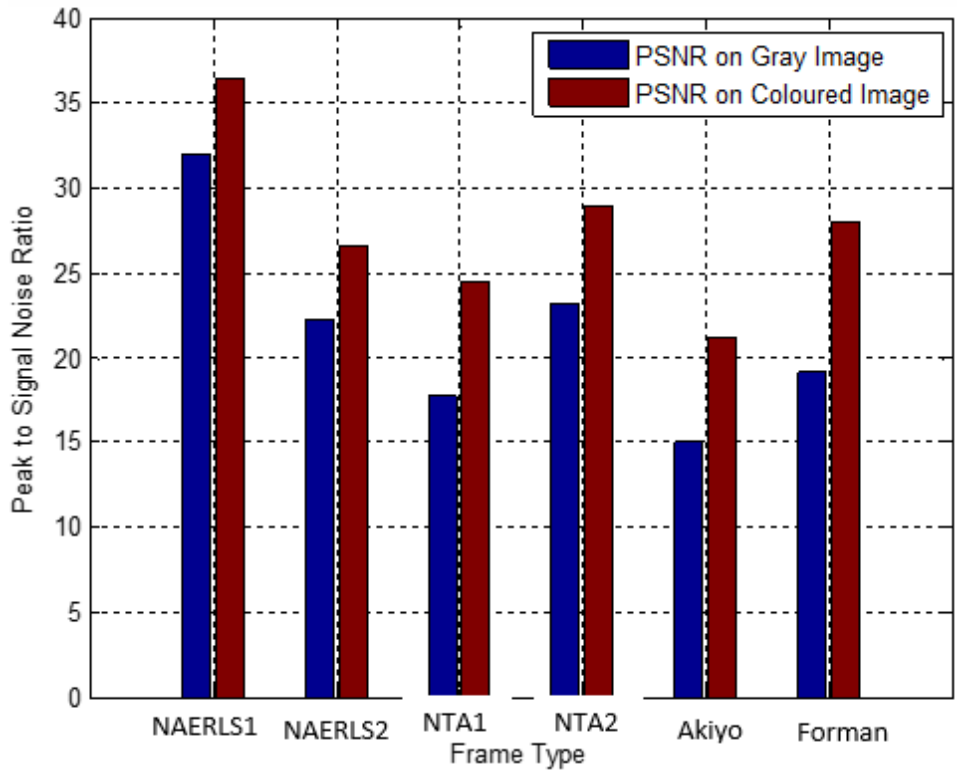

Figure 9: Brightness Enhancement Comparison on Grey scale and Coloured Frames

Figure 9 shows the amount of brightness enhancement on the grey scale frames and the coloured frames. Both *NAERLS1.avi* and *NAERLS2.avi* coloured sample frames show a PSNR percentage improvement of 12.45% and 16.32% over the grey scale sampled images. The *NTA1.avi* and *NTA2.avi* coloured frames shows a PSNR percentage improvement of 27.57% and 19.83%, respectively. Similarly, the respective PSNR percentage improvement of 28.93% and 31.68% were achieved for *Akiyo.avi* and *Forman.avi* benchmark video frames. This percentage improvement over the grey scale frames indicates the efficiency of this brightness enhancement method on coloured images. This is justifying the aim of this research work.

## 4. CONCLUSION

This research work developed a brightness enhancement technique for video frame pixel improvement based on pixel intensity analysis. A total of six (four acquired and two benchmark) sample data frames were used in the implementation and determination of the Improvement of the developed technique. Simulation results shows that the developed method is efficient with an improve pixel intensity and histogram distribution. The Peak to Signal Noise Ratio evaluation shows the efficiency and signal quality of the developed technique.